\documentclass[final,1p,times]{elsarticle}
\usepackage[T1]{fontenc}
\usepackage[latin1]{inputenc}
\usepackage{amsmath}
\usepackage{amssymb}
\usepackage{amsfonts}
\usepackage{graphicx,floatflt,epsfig}
\usepackage{accents}
\usepackage{amsthm}
\usepackage{color}
\usepackage{index}
\usepackage{bm}
\newcommand{\PF}{\psi_F}
\newcommand{\Ps}{\psi_\sigma}
\newcommand{\OF}{\Omega_F}
\newcommand{\Os}{\Omega_\sigma}
\newcommand{\vx}{\frac{\partial v}{\partial X}}
\newcommand{\Fx}{\frac{\partial F}{\partial X}}
\newcommand{\sx}{\frac{\partial \sigma}{\partial X}}
\newcommand{\eFE}{e^{(E)}_F}
\newcommand{\eVs}{e^{(V)}_\sigma}
\newcommand{\rstar}{\rho^*}

\newtheorem{remark}{Remark}
\newtheorem{theorem}{Theorem}[section]

\journal{International Journal of Non-Linear Mechanics}

\begin{document}

\begin{frontmatter}

\title{A Nonlinear approach to Viscoelasticity via \\ Rational Extended Thermodynamics}

\author{Tommaso Ruggeri} 
\ead{tommaso.ruggeri@unibo.it}

\address{Department of Mathematics, Alma Mater Research Center on Applied Mathematics (AM$^2$), \\ University of Bologna  and Accademia Nazionale dei Lincei\\
\vspace{1cm} \hspace{6cm} Dedicated to Giuseppe Saccomandi for his $60^{th}$ birthday}

\begin{abstract}
In the one-dimensional isothermal case, we introduce a simple model of nonlinear viscoelasticity within the Rational Extended Thermodynamics (RET) framework. The differential system is  determined by the universal principles of RET, exhibiting symmetric hyperbolic form and ensuring the existence of smooth solutions for appropriately small initial data. In the linear case, the equation for viscous stress reduces to the well-known Maxwell model, thereby representing a plausible nonlinear extension of the Maxwell-type model.  The total stress instead satisfies a non-linear Zener model.
\end{abstract}

\end{frontmatter}
\section{Introduction}
Viscoelasticity remains an open and fascinating topic, given its broad spectrum of potential applications. Beyond polymeric materials such as rubbers, it encompasses biological tissues like organs,  brain, skin, cardiovascular structures, connective tissues, and muscles. All soft materials, characterized by hyperelastic (nonlinear) static behavior, inherently exhibit nonlinear viscoelasticity.
The prototype for this behavior in the linear case is the Maxwell model, and various derivatives consider materials with a memory effect, i.e., non-local in time. Numerous books delve into this subject, such for example \cite{book1,book2,book3} and references therein.

It is crucial to emphasize that in continuum theories, physical laws are expressed through balance laws, necessitating constitutive equations to close the system. In the modern approach, these additional equations must adhere to universal principles like objectivity and compatibility with an entropy principle, representing the actual constitutive properties of the material.

Historically, constitutive equations were formulated mainly empirically, falling into three major classes:
\begin{itemize}
	\item  \emph{Local constitutive equations} -     
	Examples include the stress-strain relation in nonlinear elasticity and the caloric and thermal equations of state in Euler fluids, expressing internal energy and pressure as functions of mass density and temperature. Introducing such equations in the balance laws results in a hyperbolic differential system.
	\item  \emph{Non-local in space} - 
	Examples include Fourier's law for heat flux, Navier-Stokes' law for viscous stress, Fick's law for mixture diffusion, and Darcy's law for porous media. Introducing these equations in the balance laws results in a system of differential equations where some spatial derivatives are of second order, and the time derivatives are of first order. These differential systems have then a parabolic structure.
	\item  \emph{Non-local in time} -    
	Examples are viscoelastic materials, where stress depends not only on present deformation but also on its history. Except for an exponential memory kernel, the mathematical structure of such systems is of integro-differential type.
\end{itemize}

Taking into account the long-standing debate in the literature following   M\"uller seminal work  \cite{mullerfourier} which demonstrated that Fourier and Navier-Stokes constitutive equations violate objectivity, Ruggeri addressed this issue in a provocative paper titled \emph{Can Constitutive Relations be Represented by Non-local Equations?} \cite{Ruggeri_Can}. Building upon established results from RET \cite{RET,beyond,newbook}, Ruggeri argued that non-local theories in space are approximations of extended balance laws of hyperbolic type with local constitutive equations, reducing to traditional parabolic equations via Maxwellian Iteration introduced by Ikenberry and Truesdell, which is a sort of Chapman-Enskog procedure in the macroscopic case \cite{Ikenberry-1956}. 
In particular, it was shown that:

i) Navier Stokes' and Fourier's laws are a limit case of new hyperbolic balance equations for shear viscous tensor, dynamic pressure, and heat flux in RET;

ii) Fick's law is a limit case of momentum equations of each component in a mixture;

iii)  Darcy's law for porous material is a limit case of the momentum equation when inertial terms are neglected.

Therefore, the conjecture in \cite{Ruggeri_Can}  was that  \emph{true} physical systems are of hyperbolic type, which in particular is mandatory in the relativistic framework by principle.

All the previous considerations made in \cite{Ruggeri_Can} do not include the case of non-local in-time materials in which the history of the material gives the constitutive equations.

In this paper, we want to discuss some topics in the case of non-locality in time and reconsider an old idea just outlined in Chapter $16$ of the book by M\"uller and Ruggeri \cite{RET} and on the Liu's paper \cite{Liu}. The goal is to  propose a possible new approach to nonlinear viscoelasticity  using the RET postulate, i.e. requiring:

\begin{itemize}
	\item If new fluxes exist, new extended balance laws with local constitutive equations are established; 
	\item The system must be closed, compatible with the universal principles of RET: entropy principle, objectivity, and convexity of the entropy so that the system is symmetric hyperbolic.
\end{itemize}

For simplicity, we consider the one-space dimension and the isothermal case. We can prove that the universal principles are enough to determine a closed model of viscoelasticity that, in the linear case, reduces what concerns the viscous stress to the Maxwell model, while the total stress sum of the elastic part and the viscous one becomes a Zener model.

\section{The model via Rational Extended Thermodynamics}
An  elastic material in the isothermal  and one-dimension case  is  governed in  Lagrangian variables   by the following  first-order system
\begin{align}\label{elast}
	\begin{split}
		&\rho^* \frac{\partial v}{\partial t}- \frac{\partial T(F)}{\partial X} = \rho^* b, \\
		& \frac{\partial F}{\partial t}- \frac{\partial v}{\partial X}=0,
	\end{split}
\end{align}
where $\rho^*, v, T, F, b$ are the constant mass density in the reference frame, the velocity, the first Piola-Kirchhoff stress tensor, the gradient of deformation, and the external body force. In the 1-dim, all quantities are scalars and we have
\begin{equation}\label{FF6}
	F=\frac{\partial x}{\partial X}= 1+\frac{\partial u}{\partial X}, \quad v= \frac{\partial u}{\partial t}
\end{equation}
where $u = x(X,t)-X$ is the displacement.

As is well known, if we require that any solutions of \eqref{elast} are also the solutions of the supplementary energy equation (that has the role of "entropy principle" in the isothermal case) 
\begin{align*}
	\begin{split}
		&\rho^* \frac{\partial }{\partial t}\left(\frac{v^2}{2} + e(F)\right)- \frac{\partial T(F) v}{\partial X} - \rho^* b \, v =\mathcal{E} \leq 0, \\
	\end{split}
\end{align*}
we obtain
\begin{equation*}
	T(F) = \rho^* e_F(F), \qquad \mathcal{E} =0,
\end{equation*}
where $e(F)$ is the internal energy and the index from now denotes differentiation with respect to the corresponding variable (see  for example, \cite{RuggeriTermo}).

In the case of viscoelasticity, we assume that the total stress is the sum of the elastic part and the viscous one: $T(F) + \sigma$.
Differently, to consider a constitutive equation for $\sigma$ as typical of hereditary materials, we employ the ideas of RET. Since we have a new flux $\sigma$, a new balance law exists to add to \eqref{elast}. Then we consider the following differential extended system of balance laws:
\begin{align}\label{elastvisco}
	\begin{split}
		&\rho^* \frac{\partial v}{\partial t} - \frac{\partial }{\partial X}(T(F) + \sigma) = \rho^* b, \\
		& \frac{\partial F}{\partial t} - \frac{\partial v}{\partial X} = 0,\\
		& \frac{\partial \psi(F,\sigma)}{\partial t} + \frac{\partial \Omega(F,\sigma)}{\partial X} = P(F,\sigma).
	\end{split}
\end{align}
Here, the new density $\psi$, the new flux $\Omega$, and the production $P$ are constitutive equations depending locally on $(F,\sigma)$ that  need to be determined, assuming the dissipation condition ("entropy principle") that any solutions $(v(X,t), F(X,t), \sigma(X,t))$ of the system \eqref{elastvisco} also satisfy the supplementary energy dissipative balance law:
\begin{align}\label{energia}
	\begin{split}
		&\rho^* \frac{\partial }{\partial t}\left(\frac{v^2}{2} + e(F,\sigma)\right) - \frac{\partial }{\partial X}((T(F) + \sigma)v) - \rho^* b \, v = \mathcal{E} \leq 0. \\
	\end{split}
\end{align}
 To restrict the local constitutive equations 
 \begin{equation*}
 \mathcal{Z} \equiv \left\{\psi(F,\sigma), \Omega(F,\sigma), P(F,\sigma), e(F,\sigma), \mathcal{E}(F,\sigma) \right\}
 \end{equation*}
 that appear in the balance laws \eqref{elastvisco} and in the supplementary law \eqref{energia}, we use the classical method to satisfy the entropy principle used also in the elastic case of the over-determined system eliminating the time derivative in the supplementary law trough the system \eqref{elastvisco}. This method is completely equivalent to the procedure of the \emph{main field}  used as the usual procedure in RET via the Ruggeri-Strumia Theorem \ref{teorRS}.
We assume as reasonable that the internal energy is the sum of an elastic part and a viscous one:
\begin{equation*}
	e(F,\sigma) = e^{(E)}(F) + e^{(V)}(\sigma),
\end{equation*}
we require that the dissipative source $P(F,\sigma)$ vanish with  $\sigma$:
\begin{equation}\label{Pcond}
	P(F,0)=0,
\end{equation}
and 
\begin{equation}\label{psigma}
	\psi_\sigma= \left(\frac{\partial \psi}{\partial \sigma}\right)_F \neq 0,
\end{equation}
in such a way, the last equation of \eqref{elastvisco} can be put in normal form with respect to the time derivative of $\sigma$:
\begin{equation}\label{normals}
	\frac{\partial \sigma}{\partial t} = - \frac{1}{\psi_\sigma}
	\left\{\PF \vx +\OF \Fx + \Os \sx - P\right\}.
\end{equation}

 Then, we can prove the following 
\begin{theorem}\label{1teor}
	A necessary and sufficient condition such that the system \eqref{elastvisco} satisfies the dissipative supplementary equation \eqref{energia} is that
	the constitutive equations satisfy the following expressions:
	\begin{align}\label{teor}
		\begin{split}
			& T(F) = \rstar \eFE(F), \\
			&   \psi(F,\sigma) = \Psi(w),\quad w = F- Z(\sigma), \quad Z(\sigma)= \int \rstar \frac{\eVs(\sigma)}{\sigma} d\sigma, \\
			& \Omega(F,\sigma) = 0, \\
			& P(F,\sigma) = \frac{\sigma}{\alpha \Psi^\prime(w)} ,  \\
			& \mathcal{E} =-\alpha P^2, \quad \alpha \equiv \alpha(F,\sigma) >0,
		\end{split}
	\end{align}
	where $\Psi$ is an arbitrary function of $w$ and $\alpha$ is an  arbitrary positive function of $(F,\sigma)$.
\end{theorem}

{\bf Proof} :
By eliminating the time derivative in \eqref{energia} through the equations \eqref{elastvisco}$_{1,2}$ and from \eqref{normals} we obtain:
\begin{align*}
	& \left(\rstar \eFE - \rstar \frac{\eVs}{\Ps}\PF - T - \sigma \right)\vx -\\
	&  \rstar \frac{\eVs}{\Ps}\OF \Fx - \rstar \frac{\eVs}{\Ps}\Os \sx+ \rstar \frac{\eVs}{\Ps} P =   \mathcal{E} \leq 0.
\end{align*}
As the inequality needs to be verified for any processes and the spatial derivatives are arbitrary and  linear, the only possible solution is
\begin{align}\label{risultati}
	\begin{split}
		& \rstar \eFE(F) - \rstar \frac{\eVs(\sigma)}{\Ps}\PF - T(F) - \sigma =0, \\
		& \OF = \Os = 0, \quad i.e. \quad \Omega = \text{constant}=0,\\
		& \rstar \frac{\eVs}{\Ps} =- \alpha(F,\sigma) P, \quad \alpha(F,\sigma) > 0,       
	\end{split}
\end{align}
and therefore 
\begin{equation*}
	\mathcal{E} = -\alpha P^2 \leq 0.
\end{equation*}
Inserting \eqref{risultati}$_3$ into \eqref{risultati}$_1$, we have
\begin{equation*}
	\rstar \eFE(F) + \alpha P \PF - T(F) - \sigma =0.
\end{equation*}
Putting $\sigma =0$ and taking into account the assumption \eqref{Pcond}, we have:
\begin{align*}
	T(F) = \rstar \eFE(F),
\end{align*}
as in the elastic case, while from  \eqref{risultati}$_1$ we have the following partial differential equation for $\psi(F,\sigma)$:
\begin{equation}\label{eqpsi}
	\Ps + Z_\sigma(\sigma) \PF = 0, \qquad Z_\sigma(\sigma)=\rstar \frac{\eVs(\sigma)}{\sigma},
\end{equation}
that has as a solution 
\begin{equation*}
	\psi(F,\sigma) = \Psi(F - Z(\sigma)),
\end{equation*}
where $\Psi(w)$ is an arbitrary function of $w=F-Z(\sigma)$, with $Z(\sigma)$ given by integration of \eqref{eqpsi}$_2$.
Substituting into \eqref{risultati}$_3$, we have
\begin{equation*}
	P =  \frac{\sigma}{\alpha\Psi^\prime(w)},
\end{equation*}
where the prime indicates the derivative with respect to $w$. We remark that by the condition \eqref{psigma} $\Psi^\prime(w) \neq 0$.

Substituting the first four expressions of \eqref{teor}, the system \eqref{elastvisco} becomes equivalent to the following system:  
\begin{align}\label{nelastvisco}
	\begin{split}
		&\rho^* \frac{\partial v}{\partial t} - \frac{\partial }{\partial X}(\rstar \eFE(F) + \sigma) = \rho^* b \\
		& \frac{\partial F}{\partial t} - \frac{\partial v}{\partial X} = 0,\\
		& \frac{\partial }{\partial t} (Z(\sigma)-F)  = -\frac{\sigma}{\alpha \Psi^{\prime 2}}.
	\end{split}
\end{align}
Taking into account the expression of $Z$ given by the \eqref{eqpsi}$_2$ and the second of \eqref{nelastvisco}, the last equation of \eqref{nelastvisco} becomes:
\begin{equation}\label{GM}
	\rstar \frac{\eVs(\sigma)}{\sigma} \frac{\partial \sigma}{\partial t} -\frac{\partial v}{\partial X} = -\frac{\sigma}{\alpha \Psi^{\prime2}},
\end{equation}
that has the form of a nonlinear Maxwell equation. $\qed$ 

\subsection{Main field, Convexity, and Symmetric form}
Now, we want to impose the other requirements of RET requiring the convexity condition under which the system is symmetric hyperbolic and the Maxwellian iteration. First, we observe that the equation \eqref{nelastvisco}$_3$ has an infinite form of balance law for the possible choice of the function $\Psi(w)$:
\begin{equation}\label{infinite}
	\frac{\partial \Psi(w)}{\partial t} = \frac{  \sigma}{\alpha \Psi^\prime}, \qquad w= F- Z(\sigma).
\end{equation}
Then 
we recall that 
the system \eqref{nelastvisco}$_{1,2}$ and \eqref{infinite} is a particular case of a general quasi-linear  system of balance laws:
\begin{equation} \label{general}
	\frac{\partial\mathbf{F}(\mathbf{u})}{\partial t}+\frac{\partial\mathbf{G}%
		(\mathbf{u})}{\partial X}=\mathbf{f(u)}%
\end{equation}
with
\begin{equation}\label{FF}
	\mathbf{F}= \begin{pmatrix}
		\rstar  v \\  F \\ -\Psi(w)
	\end{pmatrix},
	\qquad
	\mathbf{G} = \begin{pmatrix}
		-(T(F)+\sigma) \\ -v  \\ 0 
	\end{pmatrix},
	\qquad 
	\mathbf{f}  = \begin{pmatrix}
		\rstar b \\0 \\ -\frac{\sigma}{\alpha \Psi^\prime}
	\end{pmatrix},
\end{equation}
and the energy equation \eqref{energia} is, in the general case, the supplementary quasi-linear law
\begin{equation} \label{sh1} 
	\frac{\partial h(\mathbf{u})}{\partial t}+\frac{\partial g(\mathbf{u})}{\partial X}= \Sigma(\mathbf{u}),
\end{equation}
that, in our case, is given by
\begin{equation*}
	h=\rstar \left( \frac{v^2}{2} + e^{(E)}(F) + e^{(V)}(\sigma)\right) , \quad 
	g=- \left(T(F)+ \sigma  \right)v, \quad \Sigma= \rstar b v + \mathcal{E}.
\end{equation*}
For this kind of system, we recall that 
under the thermodynamic stability condition that $h$ is a convex function with respect the field of densities $\mathbf{u} \equiv \mathbf{F}$, Ruggeri and Strumia, starting from the papers by Godunov \cite{Godunov} and Boillat \cite{Boillat}, proved the following theorem.

\begin{theorem}[Ruggeri \& Strumia \cite{RS}]\label{teorRS}
The compatibility between the system of balance laws \eqref{general} and the supplementary balance law \eqref{sh1} with $h$ being a convex function of $\mathbf{u} \equiv \mathbf{F}$ implies the existence of the \emph{main field} $ \mathbf{u}^\prime $ that satisfies
\begin{equation}  \label{prod}
	dh  = \mathbf{u}^\prime \cdot d\mathbf{F}, \qquad
	dg  = \mathbf{u}^\prime \cdot d\mathbf{G}, \qquad
	\Sigma = \mathbf{u}^\prime \cdot \mathbf{f}.
\end{equation}
If we introduce the potentials defined by
\begin{equation} \label{110}
	h^{\prime }= \mathbf{u}^\prime \cdot \mathbf{F}  - h, \qquad g^{\prime }= \mathbf{u}^\prime \cdot \mathbf{G}  - g
\end{equation} 
then \eqref{prod}$_{1,2}$ can be rewritten as
\begin{equation*}
	dh^{\prime  } = \mathbf{F}  \cdot d\mathbf{u}^\prime, \qquad   dg^{\prime  } = \mathbf{G}  \cdot d\mathbf{u}^\prime.
\end{equation*}
Choosing  the components of $ \mathbf{u}^\prime $ as field variables, we have 
\begin{equation*}
	\mathbf{F}  = \frac{\partial h^{\prime}}{\partial \mathbf{u}^\prime}, \qquad 
	\mathbf{G}  = \frac{\partial g^{\prime}}{\partial \mathbf{u}^\prime},
\end{equation*}
and then the original system \eqref{general} can be rewritten in a symmetric form with Hessian matrices:
\begin{equation} \label{simm}
	\frac{\partial}{\partial t}\left( \frac{\partial h^{\prime}}{\partial \mathbf{u}^\prime } \right) + \frac{\partial}{\partial X}\left(\frac{\partial g^{\prime}}{\partial \mathbf{u}}\right)= \mathbf{f}
	\quad \iff \quad \frac{\partial^2 h^{\prime
	}}{\partial \mathbf{u}^\prime \partial \mathbf{u}^\prime}
	\frac{\partial \mathbf{u}^\prime}{\partial t} +
	\frac{\partial^2 g^{\prime
	}}{\partial \mathbf{u}^\prime \partial \mathbf{u}^\prime}
	\frac{\partial \mathbf{u}^\prime}{\partial X}  
	= \mathbf{f}.
\end{equation}
From \eqref{110} $h^{\prime}$ is the Legendre transform of $h  $ and therefore is a convex function of $\mathbf{u}^\prime$ and therefore the first  Hessian matrix in \eqref{simm}$_2$ is positive definite. Consequently, according to Friedrichs definition \cite{KOF}, the system is symmetric hyperbolic.
\end{theorem}
\bigskip

Now we can prove the following theorem that fixes uniquely the model:
\begin{theorem}
	The only system \eqref{elastvisco} that satisfies  the universal principles of RET:
	Compatibility with a dissipation principle;
	Convexity, Stability and Symmetric hyperbolic form;
	Maxwellian Iteration;
	is the following one:
	\begin{align}\label{nnelastvisco}
		\begin{split}
			&\rho^* \frac{\partial v}{\partial t} - \frac{\partial }{\partial X}(\rstar \eFE(F) + \sigma) = \rho^* b, \\
			& \frac{\partial F}{\partial t} - \frac{\partial v}{\partial X} = 0,\\
			& \frac{\partial }{\partial t} (Z(\sigma)-F)  = -\frac{\sigma}{\mu(F)}, \quad \iff  \quad  \rstar \frac{\eVs(\sigma)}{\sigma} \frac{\partial \sigma}{\partial t} -\frac{\partial v}{\partial X} = -\frac{\sigma}{\mu(F)},
		\end{split}
	\end{align} 
	with  $Z$ given in \eqref{eqpsi}$_2$. All constitutive equations are prescribed, and the system is closed, provided we know the internal elastic energy $e^{(E)}(F)$, the internal viscous energy $e^{(V)}(\sigma)$ and the viscosity coefficient $\mu(F)$ that must satisfy the following inequalities:
	\begin{equation}\label{diseg}
		e^{(E)}_{FF}(F) >0, \qquad \frac{\eVs(\sigma)}{\sigma} >0, \qquad \mu(F) \geq 0.
	\end{equation}
	The system is symmetric hyperbolic in the form \eqref{simm} choosing as variables the main field
	\begin{equation*}
		\mathbf{u}^\prime \equiv (v,T(F) + \sigma, \sigma),
	\end{equation*}
	and having potentials:
	\begin{align*}
		& h^\prime = \rstar \frac{v^2}{2} + \rstar \left(\eFE F -e^{(E)}\right)+\sigma Z(\sigma) - \rstar \eVs, \quad g^\prime= (T+\sigma)v.
	\end{align*}
	The energy production is given by
	\begin{equation*}
		\Sigma = \rstar b v + \mathcal{E}, \quad \text{with} \quad \mathcal{E} = - \frac{\sigma^2}{\mu} \leq 0.
	\end{equation*}
\end{theorem}
\noindent
{\bf Proof}: As we have verified  the conditions for exploiting the  entropy principle in the Theorem \ref{1teor}, we are sure that there exists a main field 
\begin{equation*}
	\mathbf{u}^\prime \equiv (\xi,\zeta,\eta),
\end{equation*}
that satisfy \eqref{prod} and, in fact from
 \eqref{prod} we obtain   immediately the following expression of the  main field:
\begin{equation}\label{mainf}
	\xi= v,\qquad  \zeta=T + \sigma, \qquad \eta = \frac{\sigma}{\Psi^\prime}.
\end{equation}
The convexity requires that the quadratic form is positive:
\begin{equation*}
	Q=  \frac{\partial^2 h^{\prime
	}}{\partial \mathbf{u}^\prime \partial \mathbf{u}^\prime} \delta \mathbf{u}^\prime\cdot \delta\mathbf{u}^\prime= \delta \left( \frac{\partial h^{\prime}}{\partial \mathbf{u}^\prime}\right)\cdot \delta\mathbf{u}^\prime = \delta \mathbf{F} \cdot \delta\mathbf{u}^\prime >0.
\end{equation*}
Taking into account \eqref{mainf} and \eqref{FF}, it is simple to verify that necessary and sufficient condition such that $Q>0$ for any processes is:
\begin{equation}\label{dis}
	e^{(E)}_{FF}(F) >0, \qquad \frac{\eVs(\sigma)}{\sigma} >0, \qquad \Psi^{\prime\prime}(w) =0.
\end{equation}
The first inequality is the usual convexity condition for nonlinear elasticity; the second one if we want that the expression is bounded also for $\sigma \rightarrow 0$ implies in particular:
\begin{equation}\label{posi}
	\eVs(0) =0, \qquad e^{(V)}_{\sigma \sigma}(0)>0,
\end{equation}
while the last condition requires that $\Psi^\prime$ is constant and without loss of generality, we can put equal to $1$ taking into account that the constant can incorporate in $\alpha$ in equation \eqref{GM} and therefore
$ \Psi = F- Z(\sigma)$.

Now, as usual in RET, the parabolic limit of the system can be obtained using the \emph{Maxwellian Iteration} proposed first by Ikenberry and Truesdell \cite{Ikenberry-1956}, for more details the interested reader can see \cite{newbook}. In the present case, when we neglect the time derivative in \eqref{GM}, we need to reduce to Navier-Stokes equation with viscosity coefficient $\bar{\mu}$ that in general is a function of $F$:
\begin{equation*}
	\sigma = \bar{\mu}(F) \,\frac{\partial v}{\partial x} =\mu(F) \, \frac{\partial v}{\partial X}, \qquad \text{with} \qquad \mu(F)= \frac{\bar{\mu}(F)}{F},
\end{equation*}
and therefore it is possible to identify $\alpha = \mu$ and according with the last inequality is \eqref{teor} we have $\mu(F)\geq 0$. The potentials $h^\prime$ and $g^\prime$ can be evaluated immediately using \eqref{110}, and the theorem is proved. $\qed$

We remark that in the limit of elastic materials, the main field and the potentials are reduced to the one evaluated in \cite{ActaM}  in the isothermal case.
In the nonlinear case, the deformation tensor in $1-$dim is
\begin{equation*}
	\varepsilon = \frac{\partial u}{\partial X} + \frac{1}{2}\left(\frac{\partial u}{\partial X}\right)^2,
\end{equation*}
and therefore from \eqref{FF6} we have
\begin{equation}\label{Feps}
	F=\sqrt{1+2\varepsilon}.
\end{equation}
Inserting in the last equation of \eqref{nnelastvisco}, we have the alternative form  
\begin{equation}\label{MR}
	\rstar \frac{\eVs(\sigma)}{\sigma} \frac{\partial \sigma}{\partial t} -\frac{1}{\sqrt{1+2\varepsilon}}
	\frac{\partial \varepsilon}{\partial t} = -\frac{\sigma}{\mu(\varepsilon)}.
\end{equation}
We need to observe that $\sigma$ is only the viscous part of the stress. If we wan to rewrite the equation \eqref{MR} in terms of the total stress:
\begin{equation}
S= T(F) + \sigma = \rstar \eFE(F)
+\sigma
\end{equation}
we have:
\begin{equation}\label{total}
\rstar \frac{\eVs(\sigma)}{\sigma} \frac{\partial S}{\partial t} -\frac{1}{\sqrt{1+2\varepsilon}} \left(\rho^{*2} \frac{\eVs(\sigma)}{\sigma} e^{(E)}_{FF}(F)+1 \right)   
	\frac{\partial \varepsilon}{\partial t} = -\frac{S}{\mu(F)} + \rstar \frac{\eFE(F)}{\mu(F)}
\end{equation}
with  $\sigma = S - \rstar \eFE(F)$ and $F$ related to the deformation trough   \eqref{Feps}.
\subsection{Linear case}
In the linear case $F=1+\varepsilon$, where $\varepsilon= \frac{\partial u}{\partial x}$
is the deformation tensor in 1-dim, $\mu =$ const. and  $x=X$.
Choosing the energy quadratic:
\begin{equation*}
	e^{(E)} = \frac{\nu}{2 \rstar}\varepsilon^2, \quad
	e^{(V)} = \frac{\tau}{2\rstar \mu}\sigma^2, \quad
\end{equation*}
with $\nu$ is the Young modulus and the constant $\tau$ is a relaxation time, the system \eqref{nnelastvisco} becomes linear with the Hooke law for the elastic part $T=\nu \, \varepsilon$ and the last equation become 
the Maxwell equation of linear viscoelasticity for what concerns the viscous stress:
\begin{equation*}
	\tau
	\frac{\partial \sigma}{\partial t} -\mu \frac{\partial v}{\partial x} = - \sigma, \quad \iff \quad  \tau
	\frac{\partial \sigma}{\partial t} -\mu \frac{\partial \varepsilon}{\partial t} = - \sigma.   
\end{equation*}
Therefore, the equation \eqref{MR} is a natural nonlinear extension of the Maxwell case. 
Concerning the total stress given in \eqref{total} this reduces to
\begin{equation}
    \tau \frac{\partial S}{\partial t}- ( \nu \tau +\mu) \,\frac{\partial \varepsilon}{\partial t} = - S + \nu \, \varepsilon
\end{equation}
that is a \emph{standard linear solid model} in the Maxwell representation.
Therefore, we can consider \eqref{MR} as the nonlinear Maxwell model for the viscous stress, and for what concerns the total stress, the equation \eqref{total} is a nonlinear variant of the  Zener model \cite{Zener}.

It is essential to notice that these equations come without any direct postulations by only the universal principles of Rational Extended Thermodynamics.

\subsection{Physical interpretation}
In this simple case of $1$-dimension and for classical solutions, the new balance law \eqref{nnelastvisco}$_3$ has a simple physical interpretation. In fact, if we multiply it by $\sigma$, we can rewrite this equation as:
\begin{equation}\label{eqvr}
   \rstar \frac{\partial e^{(V)}(\sigma)}{\partial t} = \sigma \frac{\partial v}{\partial X} - \frac{\sigma^2}{\mu}.
\end{equation}
This equation represents the time evolution of the viscous energy, which is partly transformed into the power of the viscous stress, and part is lost due to dissipation.

It is important to note that this consequence arises from universal principles in one space dimension, as the new balance law \eqref{elastvisco}$_3$ is, in this circumstance, a scalar equation. In three dimensions of space, the viscous stress is a tensor denoted as $\sigma_{iA}$, leading to the new balance \eqref{elastvisco}$_3$ becoming a tensorial equation. Therefore, the 3-dimensional case requires further investigation.
\section{Characteristic velocities and K-Condition}
To evaluate the characteristic velocities $\lambda$ of the hyperbolic system \eqref{general}, as usual, we can use the following  chain rule (in $1$-dim):
\begin{equation*}
	\frac{\partial \mathbf{F}}{\partial t} \rightarrow -\lambda \delta \mathbf{F}, \qquad \frac{\partial \mathbf{G}}{\partial X} \rightarrow  \delta \mathbf{G}, \qquad \mathbf{f} \rightarrow \mathbf{0},
\end{equation*}
and we obtain from \eqref{nnelastvisco} the following algebraic homogeneous characteristic system:
\begin{align}\label{algb}
	\begin{split}
		& \lambda\, \rstar \delta v + \rstar e^{(E)}_{FF}(F) \, \delta F + \delta \sigma =0, \\
		& \lambda \, \delta F + \delta v=0,\\
		& \lambda \, \rstar \frac{\eVs(\sigma)}{\sigma} \delta \sigma + \delta v =0.
	\end{split}
\end{align}
It is simple to verify that the solutions of \eqref{algb} are:
\begin{equation}\label{contact}
    \lambda= 0 \quad \text{(contact wave)}, \quad \delta v=0, \quad \delta \sigma =  -\rstar e^{(E)}_{FF}(F) \, \delta F, \quad \delta F \neq 0,
\end{equation}
and the sound velocities:
\begin{equation}\label{vel}
	\lambda = \pm \sqrt{e^{(E)}_{FF}+ \frac{\sigma}{\rho^{*2} {\eVs}}},  \quad \text{with} \quad \delta\sigma = -\frac{\sigma}{\lambda \, \rstar \eVs(\sigma)} \delta v, \quad \delta F = - \frac{1}{\lambda} \delta v, \quad \delta v \neq 0.
\end{equation}
Taking into account the inequalities \eqref{dis}, the characteristic velocities are real and the right eigenvectors are linearly independent; in agreement with the fact that every symmetric system is automatically hyperbolic.

Moreover, the system of elasticity \eqref{elast} is a \emph{principal subsystem} of the new one \eqref{nnelastvisco} when $\sigma =0$  according to the definition and properties given in \cite{Arma}. Then, it automatically satisfies the subcharacteristic condition with respect to the elastic case, as is also evident from \eqref{vel} and \eqref{posi}:
\begin{equation*}
	\lim_{\sigma\rightarrow 0}  \left( {e^{(E)}_{FF}+ \frac{\sigma}{\rho^{*2} {\eVs}}}\right) =    {e^{(E)}_{FF}+ \frac{1}{\rho^{*2} {e^{(V)}_{\sigma\sigma}(0)}}} >   {e^{(E)}_{FF}} >0.
\end{equation*}

Therefore the sound velocity in viscoelasticity is greater concerning the ones of the elastic case.

Moreover, the so-called K-condition \cite{Kawa}, applied to the system \eqref{general}, is given by the requirement that in the characteristic algebraic system $\delta \mathbf{f}|_{eq}\neq 0$ (where {\it eq} denotes equilibrium) \cite{Jie}. In this context, it is equivalent to the condition $\delta \sigma|_{\sigma=0} \neq 0$. Notably, from the present characteristic system \eqref{algb}, this condition is satisfied for both the contact wave (see \eqref{contact}) and for the sound waves, where from \eqref{vel}$_2$, we have
\[
\delta \sigma|_{\sigma=0} = - \frac{1}{\lambda \, \rstar e^{(V)}_{\sigma\sigma}(0)}\delta v \neq 0,
\]
for the inequality \eqref{posi}$_2$. Consequently, the differential system exhibits globally smooth solutions for small initial data, aligning with well-established theorems (refer to \cite{newbook} and the references therein).

\section{Conclusions}
We systematically formulate a hyperbolic model for nonlinear viscoelasticity in one space dimension, employing the universal principles of Rational Extended Thermodynamics. The resulting differential system is symmetric hyperbolic, satisfies an entropy principle with a convex entropy, and adheres to the K-condition. Consequently, global smooth solutions exist for suitable initial data. The equation governing the viscous stress, \eqref{nnelastvisco}$_3$, or equivalently \eqref{MR}, naturally extends for the viscous stress the linear Maxwell equation for viscoelasticity and represents the decay in time of the viscous energy, as given by \eqref{eqvr}. Moreover, for what concerns the total stress, the present model indicates a nonlinear version of the so-called Zener model.

This work represents a "proof of concept", and further investigation is required to ascertain the model's ability to represent realistic viscoelastic behavior accurately. In particular, a comparison is needed between the solutions of the present model and those present in the extensive literature on nonlinear viscoelasticity, primarily dedicated to constitutive equations with memory. Saccomandi and coworkers have contributed significantly to nonlinear viscoelasticity, as seen for example, in \cite{Sac1,Sac2,Sac3}, addressing aspects such as wave propagation. We aim to compare the wave propagation phenomena in which the present model, framed within hyperbolic systems, naturally exists.

From a mathematical perspective, we contend that the sole knowledge of the internal energy and the viscous coefficient determines this model entirely. The differential system is achieved without additional postulations but solely through applying RET's universal principles.

In the RET framework, the equation governing the viscous stress is not regarded as a constitutive equation but rather as a balance law. This distinction not only allows for the definition of weak solutions, the study of shock waves, and the extension of solutions beyond the point where classical solutions may break down, but it also eliminates the necessity to satisfy the objectivity principle. Instead, only Galilean invariance is required, aligning with the nature of all balance laws in the classical framework from a physical perspective.

Generalization to the three-dimensional and/or non-isothermal case seems possible and will be addressed in a subsequent article.

\section*{Acknowledgments}
The work has been carried out in the framework of the activities of the Italian National Group of Mathematical Physics of the Italian National Institute of High Mathematics GNFM/INdAM.


\begin{thebibliography}{999}
\bibitem{book1}
R. M. Christensen, \emph{Theory of Viscoelasticity: An Introduction}, 2nd edition. Reprinted by Dover, Mineola, NY, USA (1982).

\bibitem{book2} A. S. Wineman, K. R. Rajagopal, \emph{Mechanical Response of Polymers. An Introduction}. Cambridge University Press, New York, USA (2000).

\bibitem{book3} M.  Amabili, \emph{Nonlinear Mechanics of Shells and Plates in Composite, Soft and Biological Materials},  Cambridge University Press, (2018).


\bibitem{mullerfourier} 
I. M\"uller, On the frame dependence of stress and heat flux. Arch. Rational Mech. Anal. \textbf{45}, 241 (1972). 

\bibitem{Ruggeri_Can} 
T. Ruggeri, Can constitutive relations be represented by non-local equations? Quart. Appl. Math. \textbf{70}, 597 (2012). 

\bibitem{RET} I. M\"uller, T. Ruggeri, {\em Rational Extended Thermodynamics}, 2nd ed.;  Springer: New York, NY, USA, (1998).

\bibitem{beyond} T. Ruggeri, M. Sugiyama {\em Rational Extended Thermodynamics beyond the Monatomic Gas}, Springer, Cham, Heidelberg, New York, Dordrecht, London (2015).

\bibitem{newbook} T. Ruggeri, M. Sugiyama, {\em Classical and Relativistic Rational Extended Thermodynamics of Gases}; Springer, Cham, (2021).

\bibitem{Ikenberry-1956} 
E. Ikenberry, C. Truesdell, On the pressure and the flux of energy in a gas according to Maxwell's kinetic theory. J. Rational Mech. Anal. \textbf{5}, 1 (1956).

\bibitem{Liu} I-Shih Liu, Extended thermodynamics of viscoelastic materials,
Continuum Mech. Thermodyn. {\bf 1},  143-164 (1989).

 \bibitem{RuggeriTermo} T. Ruggeri, {\em Introduzione alla Termomeccanica dei Continui}, Monduzzi Editore (2013).

\bibitem{Godunov} S. K. Godunov,  An interesting class of quasi-linear systems.  Sov. Math. Dokl.  {\bf 2}, 947--949, (1961).

\bibitem{Boillat}  G. Boillat, Sur l'existence et la recherche d'{\' e}quations de conservation suppl{\' e}mentaires pour les syst{\' e}mes hyperboliques. C. R. Acad. Sci. Paris A \textbf{278}, 909--912 (1974).

\bibitem{RS} T. Ruggeri, A. Strumia,   Main field and convex covariant density for quasi-linear hyperbolic systems: Relativistic fluid dynamics.   Ann. l'IHP Sec. A, {\bf 34}, 65--84 (1981).

\bibitem{KOF} K.O. Fredrichs, Symmetric positive linear differential equations, Comm. Pure Appl. Math. {\bf 11}, 333-418, (1958).

\bibitem{ActaM} G. Boillat, T. Ruggeri,
Symmetric form of nonlinear mechanics equations and entropy growth across a shock. Acta Mechanica \textbf{35},  271 - 274, (1980).

\bibitem{Zener} C. Zener, {\em Elasticity and Anelasticity of Metals}, University of Chicago Press, Chicago. (1948)


\bibitem{Arma} G. Boillat, T. Ruggeri, Hyperbolic principal subsystems: entropy convexity and subcharacteristic conditions, Arch. Rat. Mech. Anal., {\bf 137}, 305--320, (1997).

\bibitem{Kawa}  
Y. Shizuta, S. Kawashima, Systems of equations of hyperbolic-parabolic type with applications to the discrete Boltzmann equation. Hokkaido Math. J.  {\bf 14}, 249-275, (1985).

\bibitem{Jie}  
J. Lou, T. Ruggeri, Acceleration waves and weak Shizuta-Kawashima condition. Suppl. Rend. Circ. Mat. Palermo \textbf{78}, 187-200, (2006).

\bibitem{Sac1} L. Filograna, M. Racioppi, G. Saccomandi, I. Sgura, A simple model of nonlinear viscoelasticity taking into account stress relaxation. Acta Mechanica \textbf{204}, 21 - 36. (2009). 

\bibitem{Sac2} M. Destrade, R. W. Ogden, G. Saccomandi, Small amplitude waves and stability for a prestressed viscoelastic solid,  Zeitschrift f\"ur Angewandte Mathematik und Physik (ZAMP) {\bf 60}, 511 - 528, (2009). 

\bibitem{Sac3} H. Berjamin, M. Destrade, G. Saccomandi, Singular Travelling Waves in Soft Viscoelastic Solids of Rate Type European Journal of Mechanics/A 103 \#105144 (2024). 



\end{thebibliography}
\end{document}